# Spin lock composite and shaped pulses for efficient and robust pumping of dark states in magnetic resonance

Thomas Theis[‡], Yesu Feng[‡], Tung-Lin Wu and Warren S. Warren*
*Department of Chemistry, Duke University, Durham, NC 27708, USA.*

*warren.warren@duke.edu
[‡]These authors contributed equally.

Long-lived (symmetry protected) hyperpolarized spin states offer important new opportunities (for example, in clinical MR imaging), but existing methods for producing these states are limited by either excess energy dissipation or high sensitivity to inhomogeneities. We extend recent work on continuous-wave irradiation of nearly-equivalent spins (spin-lock induced crossing) by designing composite pulse and adiabatic shaped-pulse excitations which overcome the limitations. These composite and adiabatic pulses differ drastically from the traditional solutions in two-level systems. We also show this works in chemically equivalent spin pairs, which has the advantage of allowing for polarization transfer from and to remote spins. The approach is broadly applicable to systems where varying excitation strength induces an avoided crossing to a dark state, and thus to many other spectroscopic regimes.

Hyperpolarization methods produce nuclear magnetization many orders of magnitude larger than what is available at thermal equilibrium, and are particularly promising in clinical and preclinical applications of magnetic resonance imaging[1-3]. However, a fundamental challenge is the nuclear spin-lattice relaxation time $T_1$, which typically is too short in solution or tissue (tens of seconds for carbon-13) to monitor many meaningful biological processes. For this reason, symmetry-protected nuclear spin states (such as the singlet $S = (|\alpha\beta\rangle - |\beta\alpha\rangle)/\sqrt{2}$, which is a "dark state" with no dipole allowed transitions) have drawn considerable attention[4-17]. The first demonstrations [6, 7] used inequivalent spins to convert population from the normally accessible triplet state $T_0 = (|\alpha\beta\rangle + |\beta\alpha\rangle)/\sqrt{2}$ into the singlet, then strong spin locking or translation to a low field to preserve the singlet state. More recent work has shown that chemically equivalent [4, 18-20] or nearly equivalent spins [8, 10] can sustain long-lived states, which can be accessed by chemical transformation[4], field cycling[18], or even pulse sequences[8, 10, 19, 20]. Specifically, the so-called "M2S" sequence, consisting of precisely spaced π pulses, can interconvert magnetization and singlet-state polarization. It has become clear that multiple families of biologically compatible molecules exist that can bear protected singlet states with lifetimes of many minutes to hours, giving this approach transformative potential.

However, serious obstacles remain to using such reagents in MRI. The most important challenge is that in clinical applications, allowable energy deposition is limited. Recently, DeVience et. al.[21, 22] introduced a new approach for pumping singlet states in nearly-equivalent spins, called spin-lock induced crossing (SLIC), which drastically decreases power dissipation but is not robust to the inevitable rf or static field inhomogeneities in MRI. Here we extend their approach to the equivalent-spin case, and create an energy-efficient and robust method for population transfer using novel composite and shaped pulses. Composite pulses have been used for decades to improve robustness in magnetic resonance[23, 24] and laser[25, 26] applications, as have shaped pulses; [27-30] however, *existing shapes and solutions fail completely in this problem*, because the effective Hamiltonian is different. Our approach also has applications to a broad range of problems in coherent spectroscopy, including dressed state excitation, $T_{1\rho}$ measurements, and decoherence reduction.

Concerns about energy deposition (expressed as SAR, or the specific absorption rate) have probably dampened enthusiasm for long-lived states in the clinical community. For example, reference[6] used a locking field strength $\omega_1 = \gamma B_1 = 22,000$ rad s$^{-1}$. Current IEC normal operating limits for humans are 1.5 W/kg averaged over any 15-min period for whole body SAR[31]. SAR depends on many factors, but data from reference [32] on a head coil at 7 Tesla (300 MHz proton) with linear **B₁** lets us estimate that such a field applied for 100 s would dissipate 10,000 times the SAR limit. The newer approaches use states that are naturally immune to evolution (and thus avoid long spin locking); still, the M2S sequence in reference [20] used 96 40-μs composite π pulses ($\omega_1 = 2\pi*25000$ Hz), which under the same assumptions would dissipate 20 times the SAR limit.

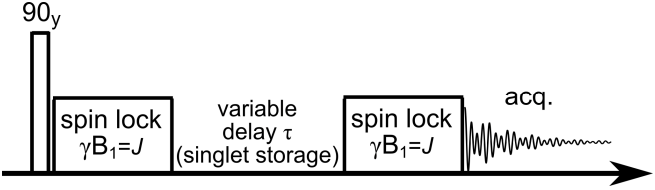

FIG. 1. Pulse sequence for the SLIC singlet-state-lifetime measurement.

In contrast, the SLIC pulse sequence in Fig. 1 [21, 22] consists of three pulses. First a hard 90-degree pulse creates transverse magnetization ($I_{1x}+I_{2x}$ in the case of a y-pulse), then the transverse magnetization is irradiated along the direction of the magnetization. A typical application requires irradiation with a locking field equal to the carbon-carbon $J$-coupling ($\omega_1 \approx 2\pi*180$ Hz) for typical times of about 250 ms, and decreases the power dissipation to <0.1 times the SAR limit.

Here we generalize SLIC to chemically equivalent spin systems that use out-of-pair spin couplings to induce singlet state polarization A simple example is $^{13}C_2$-diacetylene, H-$^{12}$C≡$^{13}$C-$^{13}$C≡$^{12}$C-H, where the carbon-12 atoms are spinless and can be ignored. This constitutes an AA'BB' system in conventional NMR notation. The two hydrogens and two carbons have the same chemical shift making them chemically equivalent, but there are two different carbon-hydrogen couplings breaking the magnetic equivalence. In papers describing the dynamics of the singlet state for a single spin pair, the singlet-triplet basis (eigenstates of $I_{1z}+I_{2z}$) is typically the most convenient basis:

$$S \equiv \frac{1}{\sqrt{2}}(|\alpha\beta\rangle - |\beta\alpha\rangle), T_1 = |\alpha\alpha\rangle, T_0 = \frac{1}{\sqrt{2}}(|\alpha\beta\rangle + |\beta\alpha\rangle), T_{-1} = |\beta\beta\rangle. \quad (1)$$

For spin locking with irradiation of the A spins, it is more convenient to switch to eigenstates of $I_{Ax}+I_{A'x}$ for A (as is also done in ref.[21, 22]) and eigenstates of $I_{Bz}+I_{B'z}$ for B:

$$S^A \equiv \frac{1}{\sqrt{2}}(|\alpha\beta\rangle_A - |\beta\alpha\rangle_A), S^B \equiv \frac{1}{\sqrt{2}}(|\alpha\beta\rangle_B - |\beta\alpha\rangle_B),$$

$$X_1^A = \frac{1}{2}(|\alpha\alpha\rangle_A + |\beta\beta\rangle_A + |\alpha\beta\rangle_A + |\beta\alpha\rangle_A),$$

$$X_0^A = \frac{1}{\sqrt{2}}(|\alpha\alpha\rangle_A - |\beta\beta\rangle_A), \quad (2)$$

$$X_{-1}^A = \frac{1}{2}(|\alpha\alpha\rangle_A + |\beta\beta\rangle_A - |\alpha\beta\rangle_A - |\beta\alpha\rangle_A),$$

$$T_1^B = |\alpha\alpha\rangle_B, T_0^B = \frac{1}{\sqrt{2}}(|\alpha\beta\rangle_B + |\beta\alpha\rangle_B), T_{-1}^B = |\beta\beta\rangle_B.$$

The most protected of the 16 states, the singlet-singlet $S_A S_B$, is symmetric with respect to interchange $(A \leftrightarrow A', B \leftrightarrow B')$, as are the nine different combinations of $X^A T^B$. Calculation of the matrix elements simplifies the problem to four relevant levels. The Hamiltonian for the spin system $H$, including the irradiated rf field amplitude $\omega_1 = -\gamma B_1$ along the x axis, is:

$$H = \begin{array}{c} \\ S^A S^B \\ X_1^A T_0^B \\ X_0^A T_0^B \\ X_{-1}^A T_0^B \end{array} \begin{pmatrix} S^A S^B & X_1^A T_0^B & X_0^A T_0^B & X_{-1}^A T_0^B \\ -2\pi J_\Sigma & \pi\Delta/\sqrt{2} & 0 & \pi\Delta/\sqrt{2} \\ \pi\Delta/\sqrt{2} & +\omega_1 & \Omega/(2\sqrt{2}) & 0 \\ 0 & \Omega/(2\sqrt{2}) & 0 & \Omega/(2\sqrt{2}) \\ \pi\Delta/\sqrt{2} & 0 & \Omega/(2\sqrt{2}) & -\omega_1 \end{pmatrix}. \quad (3)$$

Here $J_\Sigma = J_{AA'} + J_{BB'}$ is the sum of the $J$-coupling within the spin pairs and $\Delta = (J_{AB} - J_{AB'})$ reflects the out-of-pair $J$-coupling difference. By inspection the $S^A S^B \leftrightarrow X_{-1}^A T_0$ transition is put into resonance when $\omega_1 = 2\pi(J_{AA'} + J_{BB'})$. Equation (3) also makes the analogy to the optical case transparent: the irradiating field induces an AC Stark effect (raising and lowering the 1 and -1 states respectively), and the state $S^A S^B$ is a "dark state" that is made accessible by the coupling.

Near resonance ($(\omega_1 \sim 2\pi(J_{AA'} + J_{BB'}) \gg \Delta, \Omega$) the problem reduces to a two-level system, $S^A S^B$ and $X_{-1}^A T_0^B$. The (reduced) density matrix after the initial $90_y$ pulse is proportional to $\rho_1 = I_{Ax} + I_{A'x}$;

$$\rho_1 = \begin{array}{c} \\ S^A S^B \\ X_{-1}^A T_0^B \end{array} \begin{pmatrix} S^A S^B & X_{-1}^A T_0^B \\ 0 & 0 \\ 0 & -1 \end{pmatrix}. \quad (4)$$

In the SLIC experiment, rf-irradiation of amplitude $\omega_1 = 2\pi J_\Sigma$ equalizes the diagonal elements in equation [3] for $S^A S^B$ and $X_{-1}^A T_0^B$. The off diagonal element $\Delta/2\sqrt{2}$ now creates singlet population at a frequency of $\Delta/\sqrt{2}$, such that the density matrix under SLIC develops as:

$$\rho(t) = \begin{pmatrix} -\frac{1}{2} + \frac{1}{2}\cos(\Delta\pi\sqrt{2}t) & \frac{i}{2}\sin(\Delta\pi\sqrt{2}t) \\ -\frac{i}{2}\sin(\Delta\pi\sqrt{2}t) & -\frac{1}{2} - \frac{1}{2}\cos(\Delta\pi\sqrt{2}t) \end{pmatrix}. \quad (5)$$

It can also be shown that with the right combinations of couplings, SLIC works for more complicated systems such as AA'$B_2 B'_2$ with either A or B irradiation.

Figure 2 presents SLIC data for $^{13}C_2$-diphenylacetylene ($^{13}C_2$-DPA) for which we have recently reported singlet state lifetimes approaching 5 minutes using M2S[33]. The relevant parameters present in $^{13}C_2$-DPA are the $^{13}$C-$^{13}$C $J$-coupling $J_{AA'}$=182 Hz and the out-of-pair J-coupling difference $\Delta$ = 6.1 Hz. Figure 2 also shows that irradiation on the protons results in a signal enhancement of $\gamma_{1H}/\gamma_{13C} \approx 4$. With this type of spin-system it appears promising to combine the long-lived character of carbon singlets with the enhanced sensitivity of protons, thereby also improving compatibility with existing imagers that only have proton channels.

For measurements shown in Fig.2, in the absence of frequency offsets or $B_1$ inhomogeneity the expected signal intensities should be about 2.5 times larger than observed for both the "$^{13}$C in $^{13}$C out" and the "$^1$H in $^{13}$C out" measurements. The observed limitation is primarily due to $B_1$ inhomogeneity.

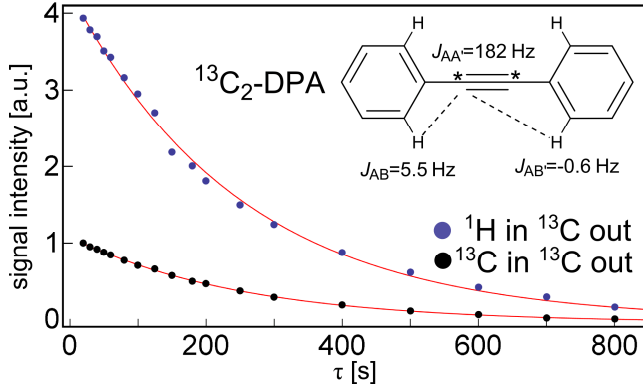

FIG 2. Singlet state lifetime measurements conducted with SLIC at 16.44 T. The data is fit to decaying exponentials with a time constant $T_S$ of 250s. Applying the first part of the pulse sequence, before the variable delay, $\tau$, on the out-of-pair protons the signal is enhanced by $\gamma_{1H}/\gamma_{13C} \approx 4$.

Frequency offsets play a secondary role here because of the high field homogeneity of the employed magnet, but would become significant in imaging applications. To gain insight into the behavior of SLIC under imperfect $B_1$ amplitude and rf-offset we simulated (Figure 3) the behavior of an AA'BB' system that approximately matches the $J$-coupling network of $^{13}C_2$-DPA ($J_\Sigma$=180 Hz and $\Delta$=5.06 Hz). For the basic SLIC experiment (Fig. 3 a-c) full singlet-triplet inversion is achieved as expected for 140 ms irradiation (= $1/(\Delta\sqrt{2})$) with $\omega_1=2\pi J_\Sigma$ on resonance. As can be seen from Figures 3b-c, the $B_1$-amplitude resonance condition is 5 Hz wide at FWHM, and the $B_1$-frequency resonance condition has a width of 62 Hz at FWHM. These values should be contrasted with typical 10% $B_1$ inhomogeneity (18 Hz) and 0.3 ppm static inhomogeneity (90 Hz in a 300 MHz imaging magnet).

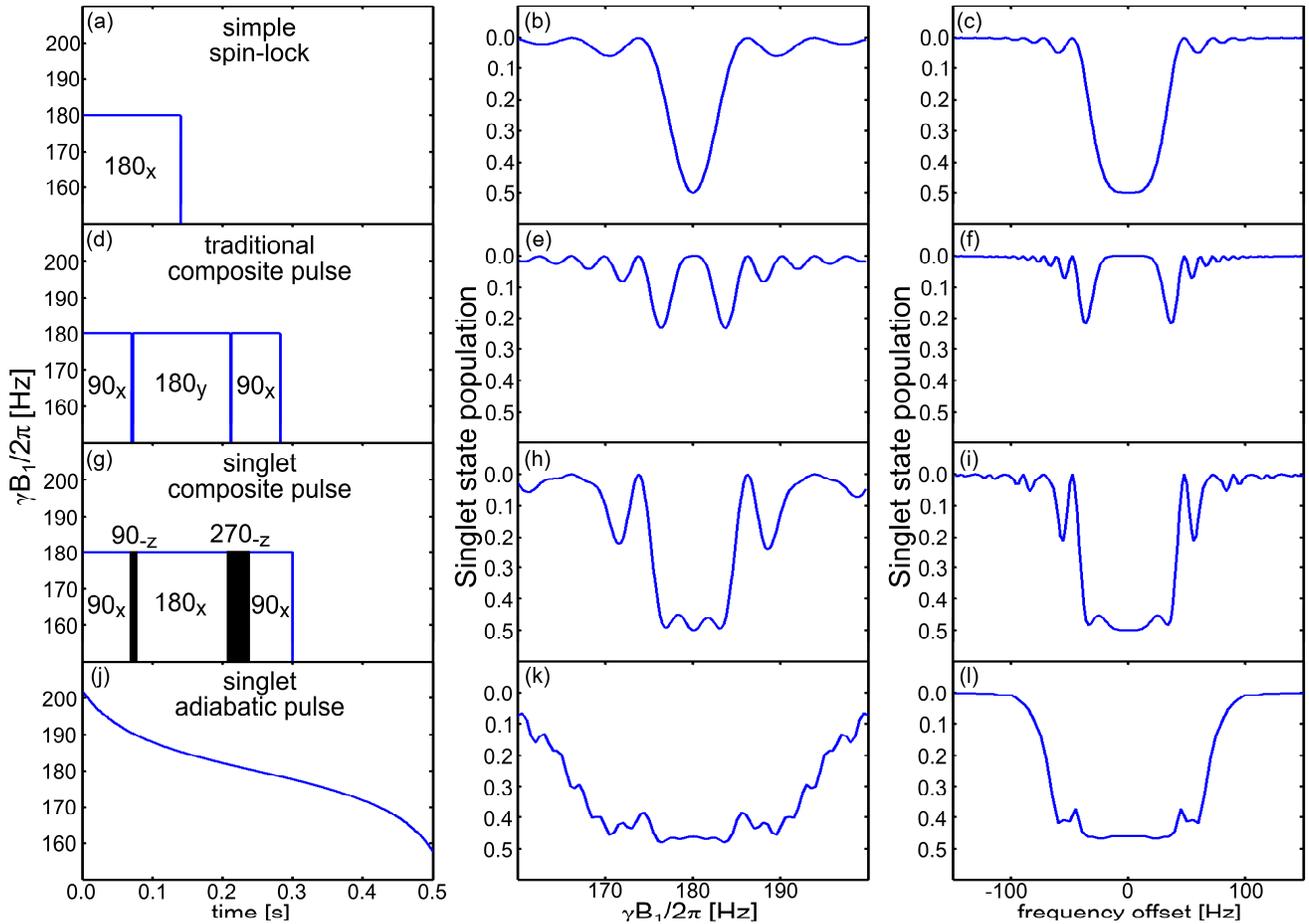

FIG. 3. Simulations of several SLIC pulse shapes examining their B1 amplitude sensitivity and frequency offset sensitivity. (Top row a-c) shows the behavior for the simple spin-lock with 5Hz bandwidth in the $B_1$ amplitude and 62 Hz bandwidth in the rf-offset. The traditional composite pulse (second row d-f) completely fails to produce the singlet state whereas singlet composite pulse (third row g-i) improves both the $B_1$-amplitude resonance condition bandwidth (10Hz) as well as the rf-offset resonance bandwidth (84Hz),. Further improvement is provided by the singlet adiabatic pulse (bottom row j-l), yielding a $B_1$-resonance bandwidth of 30Hz and rf-offset bandwidth of 140Hz.

The conventional composite pulse $90_x180_y90_x$ generates population inversion which is less sensitive to rf or static inhomogeneity than for a simple $180_x$ pulse. In fact, for just this reason, they were incorporated into the M2S sequence in many recent papers. However, as Figure 3 (d-f) shows, this approach fails completely for the SLIC sequence. The reason is that in the conventional two-level problem, the irradiating field creates an off-diagonal term with phase dictated by the pulse phase, and the diagonal term is the offset from exact resonance; but for SLIC, the irradiating field produces a *diagonal* term in the relevant basis, and the off-diagonal term is the coupling difference between the two spins. Furthermore, phase shifting the rf changes the system drastically: a 180° phase shift couples $S^AS^B$ to $X_1^AT_0^B$ instead of $X_{-1}^AT_0^B$, and a 90° phase shift couples $X_1^AT_0^B$, $X_0^AT_0^B$ and $X_{-1}^AT_0^B$. Thus, adapting shaped or composite pulses to this problem requires some transformations. The coupling $\Delta$ is constant and real, so the simplest solutions start with constant-amplitude pulses of constant phase. First we will consider the composite pulse case, and rewrite the composite pulse $90_x180_y90_x$ in an equivalent form, $90_x90_{-z}180_x270_{-z}90_x$. There are two ways to perform the z rotations. The first and simplest is to turn off the irradiation field, in which case the states are separated by $2\pi J_\Sigma$ and free evolution is equivalent to a $-z$ rotation. Thus, a free evolution of $\pi J_\Sigma/2$ gives the 90° rotation, and evolution of $3\pi J_\Sigma/2$ gives a 270° rotation. The sequence $90_x$ (free evolution, $\pi J_\Sigma/2$) $180_x$ (free evolution, $3\pi J_\Sigma/2$) $90_x$, shown in Figure 3 (g-i), generates improvement. The $B_1$-amplitude resonance condition is now 10 Hz wide at FWHM (Fig. 3h), roughly doubled from 5 Hz for the basic SLIC, and the bottom of the resonance appears much flatter. The rf-offset resonance condition is now 84 Hz wide (Fig. 3i), an improvement of 35% over basic SLIC. This sequence works well because the total needed phase shift is small. More generally, however, turning off the field during a spin locking experiment is risky, particularly if there is a range of resonance frequencies. This leads to a second and more general solution, which is to mismatch the irradiation amplitude from the exact spin locking condition. For example, a $2\pi$ pulse ($T=2\pi/\omega_1$) of *any* amplitude significantly attenuates the effect of the $\Delta$ matrix elements, because they connect states which differ in energy by $\omega_1-2\pi J_\Sigma$, but if $\omega_1 \neq 2\pi J$ such a pulse generates a z phase shift of $(2\pi(1-2\pi J_\Sigma/\omega_1))$. Thus for example, a $2\pi$ pulse with 75% of the correct spin-lock amplitude generates a phase shift of +90°, and a $2\pi$ pulse with 125% of the correct spin-lock amplitude generates a phase shift of -90°, while still being sufficiently strong to keep the two-level approximation valid.

Next we focus our attention on the implementation of adiabatic SLIC pulses for the singlet-triplet interconversion. An intuitive approach to adiabatic pulses is to switch into the frequency-modulated rotating frame introduced in Ref. [34]. That paper designs the modulated inversion pulse, with constant amplitude and a tangent frequency sweep $\Omega(t)$. This pulse gives an identical trajectory on resonance to the sech-amplitude, tanh-swept adiabatic pulses commonly used in imaging [34-36] and demonstrated with ultrafast lasers [37, 38]. As described above, the intuitive approach for the Hamiltonian in equation (3) is to consider $\omega_1$ and $J_\Sigma$ terms as z-terms and the $\Delta$ terms as x-terms, as opposed to the classical perspective used in Ref. [34] where $B_1$ amplitudes correspond to x-terms and frequency offsets correspond to z-terms. We obtain for an adiabatic SLIC pulse:

$$\omega_1(t) = \omega_c - A\cos\gamma\tan(tA\sin\gamma) \\ -\pi/(2A\sin\gamma) < t < \pi/(2A\sin\gamma), \quad (6)$$

where $\gamma$ is an adjustable parameter ($\gamma=\pi/2$ recovers the conventional unmodulated pulse) The classical "frequency modulation" is now performed by changing the amplitude of irradiation. Additionally, the center of the pulse has an amplitude of $\omega_c=2\pi J_\Sigma$, as opposed to zero-frequency offset on resonance.

Fig. 3 (j-l) displays the effects of this type of pulse, simulated for A=10.6 Hz and $\gamma$=0.455. The sensitivity to $B_1$ inhomogeneity is drastically reduced as demonstrated by the much broader $B_1$-amplitude resonance condition of Fig. (3k) For the chosen parameters the FWHM is increased by a factor 6 to 30 Hz as compared to the basic SLIC experiment. Also, the FWHM for the frequency offset is more than doubled to 140 Hz as compared to 62 Hz for the basic SLIC pulse making the adiabatic much less sensitive to off-resonance effects as well.

Finally, we contrast the performance of the basic SLIC pulse with its composite and adiabatic variants experimentally. Fig 4 shows the signal obtained for the three types of experiments normalized by simple a $^{13}$C 90-acquire experiment (y-axis) as a function of $B_1$ amplitude.

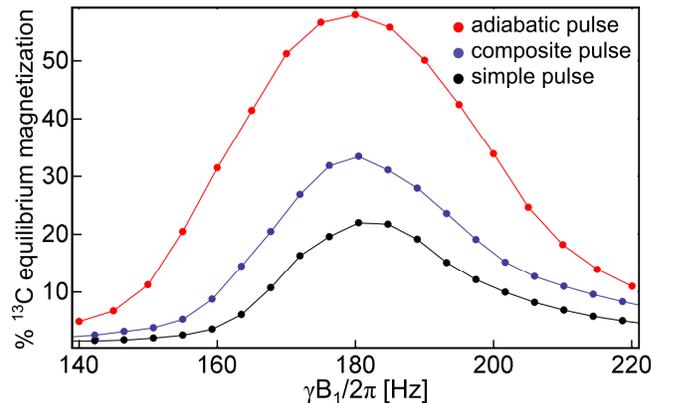

FIG. 4. Experimental evaluation of SLIC signal strength against $B_1$ field inhomogeneity. $B_1$ has been varied from 140 to 220 Hz. The composite pulse provides a 54% signal enhancement compared with simple cw irradiation whereas the adiabatic pulse gives a signal enhancement of 163%.

For these experiments a full conversion into singlet and back into triplet is performed with a relaxation delay of 5s, as depicted in Fig. 1. The first part of the sequence, creation of the singlet, is applied on the $^1$H-channel; this is also where the $B_1$ amplitude is swept. The second part, reconversion into magnetization and readout is done on resonance on the $^{13}$C-channel. With the composite pulse there is an improvement of 54% at $\omega_1=2\pi J_\Sigma$. With the adiabatic pulse the signal of the basic SLIC is outdone by 163% reaching the maximum of 60% of signal from thermal $^{13}$C-magnetization, predicted by simulations that assume neither $B_1$ amplitude inhomogeneity nor frequency offsets.

In conclusion, we have demonstrated that SLIC can be successfully implemented for chemically equivalent as well as nearly equivalent spins, and that novel composite and adiabatic versions of the SLIC pulse can improve robustness while taking advantage of the reduced power requirements of SLIC as compared to M2S. This work could be extended to amplitude-modulated shaped pulses, such as the sech-tanh pulse, by inserting unequally spaced π pulses to partially refocus the Δ coupling (at the expense of higher power dissipation) but in practice, for any given set of tradeoffs between power dissipation and needed robustness past these simple solutions, it is likely that computerized optimization is the best approach. The Hamiltonian in equation (3) has clear analogs in atomic and molecular optical spectroscopy and in general in applications where a "dark state" is made accessible by avoided crossing. Thus, this new approach is likely to have a range of practical applications.

### Acknowledgments

We thank Stephen J. DeVience, Prof. Matthew S. Rosen and Prof. Ronald L. Walsworth, for helpful discussions, most importantly those at the 54[th] Experimental NMR Conference (ENC) where their work was first presented. We also want to thank Dr. Xiaofei Liang for helping with the synthesis of $^{13}$C$_2$-diphenylacetylene.
The presented research was funded by the National Science Foundation (grant CHE-1058727)